\documentclass[aps,prl,twocolumn,superscriptaddress,showpacs,amsmath,amssymb]{revtex4-1}
\usepackage{graphicx}
\usepackage{natbib}
\usepackage{hyperref}

\newcommand{\beqa}{\begin{eqnarray}}
\newcommand{\eneqa}{\end{eqnarray}}
\newcommand{\beq}{\begin{equation}}
\newcommand{\eneq}{\end{equation}}

\begin{document}

\title{Dynamical Signatures of Edge-State Magnetism on Graphene Nanoribbons}

\author{H\'el\`ene Feldner}
\affiliation{Institut de Physique et Chimie des Mat\'eriaux de Strasbourg, UMR7504,
 CNRS-UdS, 23 rue du Loess, BP43, 67034 Strasbourg Cedex 2, France}
\affiliation{Institut f\"ur Theoretische Physik, Georg-August-Universit\"at G\"ottingen,
  Friedrich-Hund-Platz 1, 37077 G\"ottingen, Germany}
\author{Zi Yang Meng}
\affiliation{Institut f\"{u}r Theoretische Physik III, Universit\"{a}t Stuttgart, 70550 Stuttgart, Germany}
\author{Thomas C. Lang}
\affiliation{Institut f\"{u}r Theoretische Physik und Astrophysik, Universit\"{a}t W\"{u}rzburg,
  Am Hubland, 97074 W\"{u}rzburg, Germany }
  \affiliation{Institute for Theoretical Solid State Physics, RWTH Aachen University,
Otto-Blumenthal-Stra{\ss}e 26, 52056 Aachen, Germany}
\author{Fakher F. Assaad}
\affiliation{Institut f\"{u}r Theoretische Physik und Astrophysik, Universit\"{a}t W\"{u}rzburg,
  Am Hubland, 97074 W\"{u}rzburg, Germany }
\author{Stefan Wessel}
\affiliation{Institut f\"{u}r Theoretische Physik III, Universit\"{a}t Stuttgart, 70550 Stuttgart, Germany}
  \affiliation{Institute for Theoretical Solid State Physics, RWTH Aachen University,
Otto-Blumenthal-Stra{\ss}e 26, 52056 Aachen, Germany}
\author{Andreas Honecker}
\affiliation{Institut f\"ur Theoretische Physik, Georg-August-Universit\"at G\"ottingen,
  Friedrich-Hund-Platz 1, 37077 G\"ottingen, Germany}

\date{January 10, 2011; revised March 28, 2011}

\begin{abstract}
We investigate the edge-state magnetism of graphene nanoribbons using projective quantum Monte Carlo simulations and a self-consistent mean-field approximation of the Hubbard model. The static magnetic correlations are found to be short ranged. Nevertheless, the correlation length increases with the width of the ribbon such that already for ribbons of moderate widths we observe a strong trend towards mean-field-type ferromagnetic correlations at a zigzag edge. These correlations are accompanied by a dominant low-energy peak in the local spectral function and we propose that this can be used to detect edge-state magnetism by scanning tunneling microscopy. The dynamic spin structure factor at the edge of a ribbon exhibits an approximately linearly dispersing collective magnonlike mode at low energies that decays into Stoner modes beyond the energy scale where it merges into the particle-hole continuum.
\end{abstract}


\maketitle

%\section{Introduction}

Graphene is regarded as a promising candidate for future electronics. 
The electronic properties of nanodevices based on this material are in a large part governed by their edge structure.
Besides their potential for nanoelectronics and
spintronics~\cite{son06a, son06b, trauzettel07, yazyev08, rojas09},
graphene nanoribbons exhibit a remarkable interplay of low dimensionality, a
bipartite lattice and electron-electron interactions.
One of the most
fundamental predictions for graphene nanoribbons (with zigzag edges) is
the possibility of spontaneous edge-state magnetism, whereas bulk graphene is
nonmagnetic~\cite{fujita96, wakabayashi98, wakabayashi99,
rossier07, bhowmick08, jiang08, yazyev10}. 
Although electron-electron interactions are crucial for the emergence of edge-state magnetism,
essentially all these 
predictions are based on mean-field type approximations.
While such mean-field approximations were found -- by comparing to numerical exact results --
reliable at least for certain quantities and in the weak-coupling
regime~\cite{feldner10} on small honeycomb clusters,  Lieb's theorem predicts no net
magnetization for a bipartite system at half filling~\cite{lieb89} and the
Mermin-Wagner theorem \cite{mermin66} actually forbids true long-range
order for a ribbon with a fixed finite width. 
This is consistent
with a sigma-model treatment which predicts a spin gap, i.e., a finite
spin-spin correlation length for an even number of zigzag lines
\cite{yoshioka03}. Furthermore, a numerical study indeed found a
quite short correlation length in the case of two zigzag lines
\cite{hikihara03}.

In view of this puzzling situation, and given the growing experimental effort 
toward probing the magnetism of graphene nanoribbons, we believe that there
is need for an accurate numerical treatment of the Hubbard model description.
Here we present projective ground-state
quantum Monte Carlo simulations~\cite{assaad08} and compare them
with a self-consistent mean-field theory (MFT)~\cite{feldner10}.

For narrow ribbons we find indeed a finite 
correlation length; i.e., the static magnetism is an artifact of the 
mean-field approximation. However, with increasing width we observe a fast 
convergence of the qualitative behavior of the
correlation functions towards the MFT result
corresponding to strong ferromagnetic ordering tendencies. Since the 
ferromagnetic order is not static but subject to fluctuations, we next
explore the dynamical properties of the nanoribbons, focusing both on local 
and momentum-resolved single-particle and spin spectral functions. We find 
that the single-particle spectral function on the edge exhibits a 
low-energy peak as a dynamic signature of the edge-state magnetism. The 
spin spectral functions reveal the onset of magnonlike excitations on top of the  fluctuating magnetic
background along the zigzag edge. We exhibit a linear
contribution to their low-energy dispersion~\cite{wakabayashi98,You08,culchac10}
and, even more interestingly, the damping of this mode into the
Stoner continuum~\cite{culchac10}.

%\section{Statics Spin correlation along the zigzag edge}

\begin{figure}[t!]
    \centering
        \includegraphics[width=1.0\columnwidth]{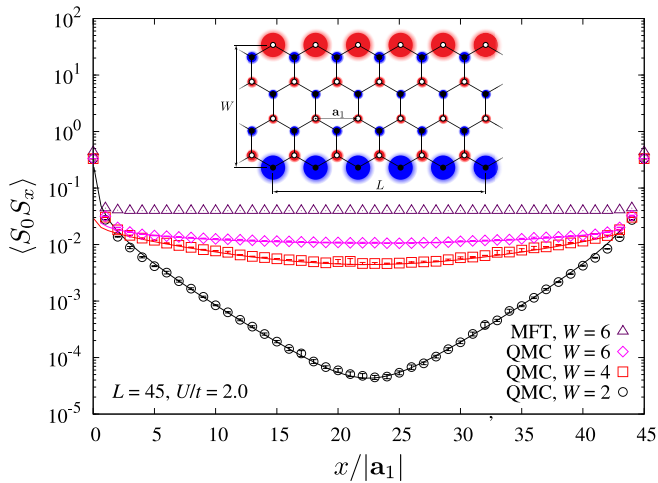}
    \caption{(Color online)
Real-space spin-spin correlation function along the zigzag edge of ribbons.
Points are the QMC and MFT data and lines are the fits with
Eq.~(\ref{eq:fittingfunction}).
Inset: Geometry of the graphene nanoribbon with zigzag edge, $L=6$, $W=4$, and
lattice size $N=2\times L\times W = 48$. The differently shaded circles represent
the positive and negative values of the local magnetization $\langle
S^{z}_{i} \rangle =\frac{1}{2}\langle n_{i,\uparrow}-n_{i,\downarrow}
\rangle$, calculated from MFT at $U/t = 2$. Radii are
proportional to the magnitude.
}
   \label{fig:correlationlength}
\end{figure}

The ribbon geometry considered here is depicted in the inset of 
Fig.~\ref{fig:correlationlength}. We employ periodic boundary 
conditions at the armchair edge and open boundary conditions at the zigzag 
edge. $L$ and $W$ denote the length and the width of the ribbon, 
respectively. $L$ is measured in units of lattice vector 
$\mathbf{a}_{1}=(\sqrt{3},0)$ and $W$ is the number of zigzag legs of the 
ribbon. The carbon-carbon bond length is set to unity.

The Hamiltonian of the Hubbard model reads
\beq
H=-t\sum_{\langle i,j\rangle ,\sigma}(c^{\dagger}_{i,\sigma}c_{j,\sigma}+c^{\dagger}_{j,\sigma}c_{i,\sigma})+U\sum_i n_{i,\uparrow}n_{i,\downarrow}
\label{Hubbard}
\eneq
with $\langle i,j \rangle$ nearest neighbors on the lattice, 
$\sigma=\uparrow,\downarrow$ and $n_i=c^{\dagger}_i c_i$.
We concentrate on half filling, i.e., the 
number of electrons is equal to the number of lattice sites.
MFT for the 2D periodic honeycomb lattice yields a 
transition from a semimetal to an antiferromagnet at $U_{c}/t=2.23$ 
\cite{MHT} whereas quantum Monte Carlo (QMC) simulations reveal a 
metal-insulator transition for $U_{c1}/t=3.5(1)$ and a further nonmagnetic 
insulator-antiferromagnetic insulator for $U_{c2}/t=4.3(1)$ \cite{meng10}. 
In this work, we focus on the interaction range $U/t \leq 2$ since we are
interested in the behavior of the system in the weak-coupling 
regime. Estimates of the local Coulomb repulsion $U$
for graphene \cite{yazyev10} lie around $U/t = 1$ and 
thus fall indeed into this regime.

The inset of Fig.~\ref{fig:correlationlength} shows the MFT result for the 
local magnetization at the edge
of a $W=4$, $L=6$ ribbon at $U/t=2$ and illustrates 
the edge-state magnetism which is expected for weak Coulomb interactions $U/t>0$ 
\cite{fujita96, wakabayashi98, wakabayashi99, rossier07, 
bhowmick08, jiang08, yazyev10}. This is reflected by the MFT result for 
the spin-spin correlation function which is shown by triangles
for $W=6$, $L=45$, and 
$U/t=2$ in the main panel of Fig.~\ref{fig:correlationlength}: the 
correlation function rapidly approaches a constant value as a function of 
distance $x$ along the edge with very little dependence on the width $W$ 
of the ribbon (not shown).

By contrast, the QMC results for the static spin-spin correlation function 
shown in the main panel of Fig.~\ref{fig:correlationlength} exhibit a 
clear decay, at least for narrow ribbons with $W=2$ (circles)
and $4$ (squares). Note that the 
rapid decay for $W=2$ is consistent with a previous density-matrix renormalization group study 
\cite{hikihara03}. However, as the width increases, the decay becomes 
slower and the QMC result for $W=6$ in Fig.~\ref{fig:correlationlength}
(diamonds) is  already qualitatively similar to the MFT result.

For a more quantitative analysis we fit the QMC data for the longest
available ribbons with a function combining exponential and power-law
behavior
\beq
%\[
\langle S_{0}S_{x}\rangle = C \left(
x^{-\eta}e^{-x/\xi}+(L|\mathbf{a}_{1}|-x)^{-\eta}\,
e^{-(L|\mathbf{a}_{1}|-x)/\xi}\right) \, .
\label{eq:fittingfunction}
%\]
\eneq
From a fit for $L=60$ at $W=2$ we estimate $\xi/|\mathbf{a}_{1}|=4.0\pm1$ and 
$\eta=0.5\pm0.3$ and at $W=4$ the $L=54$ data yield
$\xi/|\mathbf{a}_{1}|=15\pm 4$ 
and $\eta=0.2\pm0.15$. At $W=6$ it is no longer possible to distinguish 
the correlation length from infinity or the exponent $\eta$ from zero
for the largest available system size within our study ($L=45$). 
Indeed, the $W=6$ QMC data can equally well be fitted by 
predictions based on an infinite correlation length.
We thus arrive at a picture
similar to even-leg spin ladders \cite{DagRi96} or integer spin-$S$
quantum chains \cite{todo01}: ribbons with even $W$ have a finite
spin-spin correlation length \cite{yoshioka03}, but
the correlation length rapidly grows with $W$ such that for
practical purposes it can be considered infinite for $W 
\gtrsim 6$. In this sense edge-state magnetism is found
already for ribbons of moderate widths. In order to check if
MFT becomes also quantitatively accurate for wider ribbons,
one would need to go beyond the system sizes which we can presently
access by QMC calculations.

%\section{Spectral functions}

\begin{figure}[t!]
    \centering
        \includegraphics[width=1.0\columnwidth]{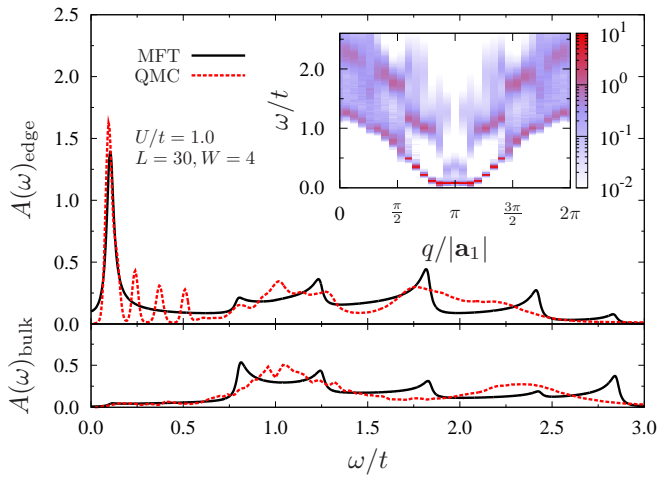}
    \caption{(Color online)
Spectral functions of a width $W=4$ ribbon at $U/t=1$.
QMC data are for a ribbon of length $L=30$ whereas MFT corresponds
to the thermodynamic limit. The inset shows QMC results for the
momentum-resolved spectral function $A(q,\omega)$ along the zigzag
edge, while the main panel shows the
local spectral function $A(\omega)$ subject to a Lorentzian broadening
$\Delta\omega = 0.02\,t$.
Note the prominent low-energy peak in  $A(\omega)_{\text{edge}}$
at $\omega/t \approx 0.1$ which is absent in $A(\omega)_{\text{bulk}}$.
}
   \label{fig:Aw-MFT-QMC}
\end{figure}

Since the ferromagnetic behavior at the edges remains fluctuating in a 
proper treatment of the Hubbard model (\ref{Hubbard}), we now turn to the 
dynamic properties. In this respect,
the local single-particle spectral function $A(\omega)$
is a particularly important observable since it can be observed in
scanning tunneling microscopy (STM) experiments \cite{niimi10,ugeda10}.
However, to the best of our knowledge, previous theoretical studies
of the interplay of edge-state magnetism and spectral functions
of graphene ribbons are restricted to density-functional
approaches \cite{jiang08,pan10} with an inherent mean-field type approximation.
In QMC we measure the 
momentum-resolved Green's function in imaginary time $G(q,\tau)$,
and then apply a 
stochastic analytical continuation scheme~\cite{sandvik98, beach04} to 
rotate $G(q,\tau)$ on the $\tau$ axis to $A(q,\omega)$ on the $\omega$ axis.
Finally, the local spectral function $A(\omega)$ is obtained by integration
over the momentum $q$ along the ribbon direction.

Figure \ref{fig:Aw-MFT-QMC} shows the single-particle spectral function
both on the zigzag edge 
$A(\omega)_{\text{edge}}$ and inside the bulk $A(\omega)_{\text{bulk}}$
for ribbon of width $W=4$ at $U/t=1$. The local spectral functions in
the main panel were subjected to a Lorentzian broadening of $\Delta\omega =0.02\,t$.
$A(\omega)_{\text{edge}}$ exhibits a dominant low-energy peak
at $\omega/t \approx 0.1$ which is absent in $A(\omega)_{\text{bulk}}$.
This peak can be traced to a flat region of a single-particle
dispersion visible in $A(q,\omega)_{\text{edge}}$ (compare the inset
in Fig.~\ref{fig:Aw-MFT-QMC}). The agreement between MFT and QMC is
excellent for this low-energy peak and the overall agreement is also
good at higher energies. The three additional sharp features
in the QMC result for $A(\omega)_{\text{edge}}$ in the region
 $0.2 < \omega/t <  0.6$ can be traced
to a finite-size momentum discretization effect and are reproduced by MFT
if the latter is also restricted to a length $L=30$ ribbon. We conclude
that the agreement between MFT and QMC for $A(\omega)$
is remarkable even at a quantitative
level, at least for weak interactions $U/t \lesssim 1$.

\begin{figure}[t!]
\centering
\includegraphics[width=\columnwidth]{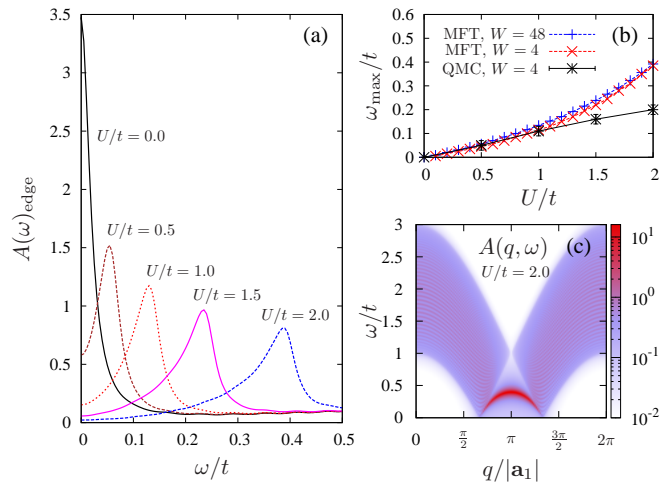}
\caption{(Color online)
(a) Single-particle spectral function 
$A(\omega)_{\text{edge}}$ for several values of $U/t$ and a ribbon of width
$W=48$. (b) Comparison of the position $\omega_{\text{max}}$ of the maximum of
$A(\omega)_{\text{edge}}$ between MFT for $W=4$, $48$ and QMC for $W=4$ and $L=30$.
(c) $A(q,\omega)_{\text{edge}}$ for a ribbon with $W=48$ at $U/t=2$.
Unless stated otherwise, results are based on MFT for long ribbons and a Lorentzian broadening
$\Delta\omega =0.02\,t$.}
   \label{fig:Aw-MFT}
\end{figure}

Having gained confidence in the accuracy of the MFT results for $A(\omega)$,
we use it to analyze a bigger
and more realistic system with $W=48$, corresponding
to a graphene nanoribbon of 10~nm width. 
The evolution of the low-energy peak on the edge
with the Coulomb repulsion is shown in Fig.~\ref{fig:Aw-MFT}(a). 
The energy $\omega_{\text{max}}$ corresponding to the maximum intensity of
the spectral function
increases with the Coulomb repulsion $U$ and is located at $\omega_{\text{max}}
=0$ only for $U=0$, i.e., the noninteracting system. In combination with
the fact that this peak exists only on the ferromagnetic edge (compare
Fig.~\ref{fig:Aw-MFT-QMC}) we conclude that it is a clear dynamic signature of
edge-state magnetism.
Figure \ref{fig:Aw-MFT}(b) demonstrates
that $\omega_{\text{max}}$ is insensitive to the actual
width of the ribbon and confirms that MFT is accurate for
$U/t \lesssim 1$ (the fact that MFT becomes less accurate as one approaches
the mean-field phase transition at  $U_{c}/t=2.23$ \cite{MHT} is not
surprising).

The momentum-resolved spectral function [shown in
Fig.~\ref{fig:Aw-MFT}(c) for $U/t=2$] reveals that the feature at
$\omega_{\text{max}}$ can be traced to the maximum of
a single-particle band at $q/|\mathbf{a}_{1}|=\pi$
where large matrix elements and a van Hove singularity combine
to yield a maximum intensity of the local spectral function. The
true single-particle gap $\Delta_{\text{sp}}$ is located in the vicinity of
$q/|\mathbf{a}_{1}|=2\,\pi/3$ and $4\,\pi/3$. For the $W=48$ ribbon shown in
Fig.~\ref{fig:Aw-MFT} it is only $\Delta_{\text{sp}}/t = 0.037$ for $U/t=2$,
resulting in a fill-in of spectral
weight in the local density of states below the ``pseudogap''
$\omega_{\text{max}}$ [compare Fig.~\ref{fig:Aw-MFT}(a)].
The value of the single-particle gap $\Delta_{\text{sp}}$ depends strongly
on the width of the ribbon; for the $W=4$ ribbon shown in
Fig.~\ref{fig:Aw-MFT-QMC}
it is quite close to the pseudogap $\omega_{\text{max}}$.
Note that the data in Fig.~\ref{fig:Aw-MFT}(a) are subjected to
a broadening $\Delta\omega =0.02\,t$. With a higher energy resolution,
one would observe further features, in particular
another van Hove singularity at $\Delta_{\text{sp}}$, albeit
with a much smaller weight than at $\omega_{\text{max}}$.

%\section{Spin spectral function}

Another probe to exhibit the peculiar magnetic properties
of zigzag nanoribbons is provided by
the spin spectral function $S(q,\omega)$ on the edge.
From the enhanced ferromagnetic correlations along the zigzag edge at finite values of $U$, 
one might expect $S(q,\omega)$  to exhibit
collective spin-wave (magnon) excitations characteristic of a ferromagnetic background,
with a quadratic low-energy dispersion relation near $q=0$. 
However, as pointed out in Refs.~\cite{wakabayashi98,culchac10},
the magnetic excitations are affected by the antiferromagnetic coupling between the two edges
of the nanoribbon that is mediated by the bulk conduction electrons. 
These antiferromagnetic correlations result in a
linear contribution to the dispersion at small wave vectors
(see also Ref.~\cite{You08}).
Furthermore, for energies entering the particle-hole continuum,
the collective modes can decay into strongly damped Stoner
modes~\cite{culchac10}. 

\begin{figure}[t!]
\centering
\includegraphics[width=1.0\columnwidth]{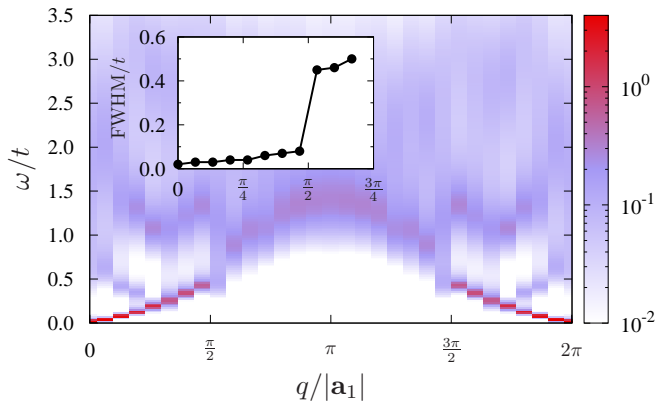}
\caption{(Color online) QMC results for the
spin spectral function $S(q,\omega)$ along the edge of a zigzag ribbon with $L = 30, W = 4$ and $U/t=2$.
The inset shows the linewidth (FWHM) of the dominant mode in $S(q,\omega)$ as a function of momentum $q$.}
\label{fig:Sw-MFT-QMC}
\end{figure}

Our QMC data in  Fig.~\ref{fig:Sw-MFT-QMC} indeed exhibit such
a scenario for a ribbon with $W=4$: an approximately linearly dispersing
sharp mode is observed up to a wave vector $q/|\mathbf{a}_{1}|\approx  0.5\,\pi$ with a 
threshold energy $\omega_{\text{th}}/t = 0.42\pm0.04$, compatible with twice  the
corresponding single-particle gap of $\Delta_{\text{sp}}/t=0.2\pm0.02$.
The inset of Fig.~\ref{fig:Sw-MFT-QMC} 
shows the corresponding evolution of the linewidth (FWHM) of this
low-energy mode that drastically increases beyond $q/|\mathbf{a}_{1}|\approx 0.5\,\pi$. 

%\section{Conclusion}

In summary, we have investigated the static as well as dynamic properties 
of graphene nanoribbons with zigzag edges.
Upon closer inspection, true
ferromagnetic long-range order at the zigzag edge
\cite{fujita96,wakabayashi98, wakabayashi99, rossier07, bhowmick08,
jiang08, yazyev10} of a finite-width ribbon turns out to be
absent \cite{lieb89,mermin66,yoshioka03,hikihara03}, yet there are
strong ferromagnetic correlations which are essentially long range but for
the narrowest ribbons. As a next step, we have identified a dominant
low-energy peak in the local spectral function as a dynamic signature
of edge-state magnetism, thus providing a theoretical guide for further
STM and spin-resolved STM experiments.
The agreement between QMC and MFT for the local spectral function
is remarkably accurate for moderate Coulomb interactions,
justifying a description of realistic geometries within a MFT framework.
In particular, we have demonstrated that the position of the dominant
low-energy peak at the zigzag edge of a nanoribbon is controlled mainly
by the Coulomb interaction $U$ such that STM experiments can be used
to deduce the appropriate value for graphene. Lastly, we have presented
QMC results for the dynamic spin structure factor at the edge
of a width $W=4$ ribbon
and demonstrated the presence of an approximately
linearly dispersing low-energy collective
spin-wave excitation \cite{wakabayashi98}
that decays upon entering the particle-hole continuum~\cite{culchac10}.

%%%%%%%%%%%%%%%%%%%%%%%%%%%%%%%%%%%%%%%%%%%%%%%%%%%%%

\begin{acknowledgments}
We wish to thank the DAAD for Grant No.\ A/10/70636 as well as the Deutsche
Forschungsgemeinschaft for financial support through SFB/TRR21, SFB602 and
Grants No.\ AS 120/4-3, No.\ HO 2325/4-2, No.\ WE 3649/2-1. Z.~Y.~M, T.~C.~L.,
F.~F.~A., and S.~W.\ acknowledge the Kavli Institute of Theoretical Physics at UCSB for hospitality through NSF Grant No. PHY05-51164. We also acknowledge NIC J\"ulich, HRLS Stuttgart, and the BW Grid for allocation of CPU time.
\end{acknowledgments}

{\it Note added:} Recently, we became
aware of Ref.~\cite{tao11} which has observed low-energy spectral
features by STM and compared those to MFT
for isolated chiral graphene nanoribbons.

%%%%%%%%%%%%%%%%%%%%%%%%%%%%%%%%%%%%%%%%%%%%%%%%%%%%%
%% bibtex screws up on "\~"
\def\Tilde#1{\~{#1}}
\bibliography{biblio-ribbon-MFT-QMC}
%%%%%%%%%%%%%%%%%%%%%%%%%%%%%%%%%%%%%%%%%%%%%%%%%%%%%

\end{document}